\documentstyle[12pt,epsf]{article}
\newcommand {\bi} {\bibitem}
 \newcommand {\be} {\begin{equation}}
\newcommand {\bea} {\begin{eqnarray} \nonumber }
\newcommand {\ee} {\end{equation}}
\newcommand {\eea} {\end{eqnarray}}
 \newcommand {\eps} {\epsilon}
 \newcommand {\si} {\sigma}
\newcommand {\de} {\delta}

 \newcommand {\al} {\alpha}

\newcommand {\ba} {\overline}
\newcommand {\lan} {\langle}
\newcommand {\ran} {\rangle}

\newcommand {\cC}  {{\cal C}}

\newcommand {\for} {\ \ \ \mbox{for}\ \ }

\def \form#1 {eq. (\ref{#1}) }
\def \parziale#1#2  {{\partial {#1} \over \partial {#2}}}

\begin{document}

\title{On the replica method for glassy systems}
\author{ Giorgio Parisi \\
         Dipartimento di Fisica and INFN, Universit\`a di Roma {\sl La Sapienza},\\
        Piazzale Aldo Moro, Roma 00185, Italy }
\maketitle

\begin{abstract}
In this talk we review our theoretical understanding of spin glasses paying a particular attention 
to the basic physical ideas.  We introduce the replica method and we describe its probabilistic 
consequences (we stress the recently discovered importance of stochastic stability).  We show that 
the replica method is not restricted to systems with quenched disorder.  We present the consequences 
on the dynamics of the system when it slows approaches equilibrium are presented: they are confirmed 
by large scale simulations, while we are still awaiting for a direct experimental verification.
\end{abstract}

\section{INTRODUCTION}

It is a sad occasion to speak at this meeting.  Giovannino was a student of mine during the first 
year I was giving a course at the university (in 1979).  It was a particular occasion for me (my 
first teaching as a professor) and I have a vivid memory of him during that course, of his positive 
attitude toward life, of his smiling.  Later Giovannino went on his road which was different but 
nearly parallel to mine.  We have written a few papers together \cite{INSIEME} and I remember with 
great pleasure this collaboration. Just a few days before his unexpected death, we spent more than 
an hour discussing his recent work on spin glasses \cite{PALA} and the possibility of extending it 
to the case of spontaneous breaking of the replica symmetry.

In this talk I will present a short introduction to the replica method for glassy system, 
concentrating on the physical ideas; I have tried to organize this in such a way that i believe it 
should have pleased Giovannino, if he were present.

In section II we introduce the replica method and we describe its probabilistic consequences.  In 
section III we define a recently discovered property (stochastic stability) and we show its great 
relevance in understanding the properties of disordered systems.  In section IV we present some of 
the considerations that imply that the replica method is not restricted to systems with quenched 
disorder.  In section V we spell out the consequences of the replica approach on the dynamics of the 
system when it slows approaches equilibrium are presented and we show that they are confirmed by 
large scale simulations.  Finally, in section VI, we present our conclusions.

\section{THE REPLICA METHOD}

The basic idea of the replica method is quite simple.  We consider a function $F(n)$ defined on the 
integers; using some property of this function, we extend it to real numbers.

As far I can tell the first 
use of the replica method goes back to Nicola d'Oresme, bishop of Lisieux (1330-1378).  He made the 
following observation.  It well known that
\be
(a^{m})^{n}= a^{mn}
\ee
for integer $n$.  If we suppose that $a^{m}$ make sense also for rational non-integer $m$, the 
previous property allow its computation.  For example using
\be
(a^{1/2})^{2}= a,
\ee
we find that 
\be
a^{1/2}= \sqrt{a}.
\ee

The same procedure can be used in the study of disordered systems \cite{mpv,PARIBOOK}.  In this case 
we have an Hamiltonian $H_{J}(\si)$ in which the $J$ are some control variables (which are 
distributed according to the probability $P(J)$) and the $\si(i)$ ($i=1,N$) are the dynamical 
variables.  We define
\be
Z_{J}=\sum_{\{\si\}}\exp (-\beta H_{J}(\si))\equiv \exp(-\beta N F_{J}).
\ee

Our goal is to compute 
\be
\ba{F}\equiv \int dP(J) F_{J}.
\ee
(the bar denotes the average of the control variables $J$).

 At this end we introduce for real $n$ the function $F(n)$ defined as
\be
F(n)= -{\ln\left(\ba{Z^{n}_{J}}\right) \over n \beta N}\ .
\ee
The value of $F(0)$ can be defined by continuity in $n$ and we find that
\be
F(0)=\ba{F}.
\ee

On the other hand for integer $n$ we can introduce $n$ replicas of the same system (which will be 
labeled by $a=1,n$) and can write
\be
\ba{Z^{n}_{J}}= \ba{\prod_{a=1,n}\left(\sum_{\{\si^{a}\}}\right) \exp(-\beta 
\sum_{a=1,n}H_{J}(\si_{a})})=
\prod_{a=1,n}\left(\sum_{\{\si^{a}\}}\right) \exp(-\beta
H_{eff}(\si))
\ee
where $H_{eff}(\si)$ depends on the $\si$-variables in all the $n$ replicas.  For example if
\be
H=\sum_{i,k=1,N}J_{i,k}\si_{i}\si_{k},
\ee
we have that
\be
H_{eff}(\si)=\sum_{a,b=1,n}\sum_{i,k=1,N}\si^{a}_{i}\si^{a}_{k}\si^{b}i_{i}\si^{b}_{k}.
\ee

There are many systems whose dynamics becomes very slow at low temperature.  Some of these systems 
clearly display a thermodynamic transition (as can be seen by many effects, e.g.  discontinuities in 
the specific heat, sharp peaks in the susceptibilities or divergences in non linear 
susceptibilities).  However by lowering the temperature we do not produce an ordered state like a 
crystal.

The aim of the replica theory is to describe this transition and the system below the transition. 
 
Let us first recall the description of these glassy systems at equilibrium in the low energy 
phase according to the predictions of the replica theory.  We consider a given system and we denote 
by $\cal C$ a generic configuration of the system.  For simplicity we will assume that there are no 
symmetry in the Hamiltonian (in presence of symmetries the arguments must be slightly modified).

It is 
useful to introduce an overlap $q(\cC,\cC')$.  There are many ways in which an overlap can be 
defined; for example in spin system we could define
\be
q={\sum_{i=1,N}\si_{i}\tau_{i}\over N},
\ee
$N$ being the total number of spins or particles and $\si$ and $\tau$ are the spins of the two 
configurations. 

 In a liquid a possible definition of the overlap is given by
\be
 q={\sum_{i=1,N}\sum_{k=1,N}f(x(i)-y(k))\over N},
\ee
where $f(x)$ is a function which decays in a fast way at large distances and is substantially 
different from zero only at distances smaller that the interatomic distance ($x$ and $y$ are the 
coordinates of the $N$ particles of the two configurations of the system).

In the high temperature phase for very large values of $N$  the probability distribution 
of the overlap ($P_{N}(q)$) is given by
\be
P_{N}(q)\approx \delta (q-q^{*}).
\ee
The value of $q^{*}$ can be often simply computed.  For example for spin systems in zero magnetic 
field we have $q^{*}=0$.  In a liquid $q^{*}=\rho\int d^{3}x f(x)$.

 In the low temperature phase $P_{N}(q)$ depends on $N$ (and on the quenched disorder, if it is 
present).  When we average over $N$ we find a function $P(q)$ which is not a simple delta function.  
In all known cases \cite{mpv,PARIBOOK,BOOK} one finds that
\be
P(q)=a_{m}\delta(q-q_{m})+a_{M}\delta(q-q_{M})+p(q),
\ee
where the function $p(q)$ does not contains delta 
function and its support is in the interval $[q_{m},q_{M}]$.

The non triviality of the function $P(q)$ (i.e.  the fact that $P(q)$ is not a single delta 
function and consequently $q$ is an intensive fluctuating quantity) is related to the existence of 
many  different equilibrium states.  Moreover the function $P_{N}(q)$ changes with $N$ and its 
statistical properties (i.e the probability of getting a given function $P_{N}(q)$) can be 
analytically computed \cite{mpv}.

In this equilibrium description a crucial role is given by the function $x(q)$ defined
as
\be
x(q)=\int_{q_{m}}^{q}P(q')dq'\ .
\ee

In the simplest case the function $P(q)$ is equal to zero. i.e the function $P(q)$ has only two delta 
functions without the smooth part.  In this case, which correspond to one step replica symmetry 
breaking, there are many equilibrium states, labeled by $\al$, and the overlaps among two generic 
configurations of the same state and of two different states are respectively equal to $q_{M}$ and 
$q_{m}$.  The probability ${\cal P}(f)$ of finding a state with total free energy $f$ is 
proportional to
\be
{\cal P}(f) \propto \exp \left( m\beta(f- f_{R})\right), \label{REM}
\ee
where $f_{R}$ is a reference free energy and $m$ is the value of $x(q)$ in the interval 
$[q_{m},q_{M}]$.

In the more complicated situation where the function $p(q)$ is non zero, couples of different states 
may have different values of the overlaps.  The conjoint probability distribution of the states and 
of the overlaps can be described by formulae similar to eq.  (\ref{REM}), but more complex 
\cite{mpv}.

\section{STOCHASTIC STABILITY}

Stochastic stability is a property which is valid in the mean field approximation; it is however 
possible to conjecture that is valid in general also for short range models.  It has been 
introduced quite recently \cite{GUERRA,AI,SOL,FMPP} and strong progresses have been done on the study 
of its consequences.

In order to decide if a system with Hamiltonian $H$ is stochastically stable, we have to consider 
the free energy of an auxiliary system having the following Hamiltonian:
\be
H+\eps^{1/2}H_{R}.
\ee
If the average (with respect to $H_{R}$) free energy is a differentiable function of $\eps$ (and the 
limit volume going to infinity commutes with the derivative with respect to $\eps$), for a generic 
choice of the random perturbation $H_{R}$ inside a given class and $\eps$ near to zero, the system is 
stochastically stable.
 
The definition of stochastic stability may depend on the class of random perturbations we 
consider.  Quite often it is convenient to chose as a random perturbation an infinite range 
Hamiltonian, e.g.
\be
H_{R}=\sum_{i,k,l}J_{i,k,l} \si_{i} \si_{k} \si_{l} \label{SSS}
\ee
where sum runs over all the $N$ points of the system and the $J$'s are random variables with 
variance $1/N$.

In the nutshell stochastic stability tell us that the Hamiltonian $H$ does not has any special 
features and that it properties are quite similar to those of similar random systems.

Although it seems quite natural, stochastic stability has quite deep consequences.  For example we 
could consider a system in which there are many equilibrium states, labeled by $\al$, and the 
overlaps among two generic configurations of the same state and of two different states are 
respectively $q_{M}$ and $q_{m}$, the free energies of the different states are uncorrelated\ldots 
The situation would be quite similar to the one described by one step replica symmetry breaking.  
However we may not specify the form of the probability distribution of the free energies which is 
characterized by a function ${\cal P}(f)$ which a priori may have an arbitrary shape.

It is a simple computation to verify that stochastic stability implies that the probability 
distribution of the free energies (${\cal P}(f)$) must have the form given in eq.  (\ref{REM}) with 
an appropriate choice of $m$.  The most dramatic effect of stochastic stability is to link the 
behaviour of the function ${\cal P}(f)$ in the region of large $f$ (where a large number of states 
do contribute) to the low $f$ behaviour, which controls the distribution of the states which are 
dominant in the partition function.

We have seen that stochastic stability strongly constraints the properties of the systems and many 
of the qualitative results of the replica approach can be derived as mere consequences of stochastic 
stability.  Stochastic stability apparently does not imply ultrametricity, which seems to be an 
independent property \cite{SOL}.  This independence  problem is still open as far has the only 
explicitly constructed probabilities distribution of the free energies of the states are 
ultrametric.

\section{SYSTEMS WITOUT QUENCHED DIS\-ORDER}
Apparently the previous discussions are restricted to systems where  quenched disorder is present.  
The requirement of quenched disorder would limit ourselves in the applications of the replica method 
and one would cut all those systems, like glasses, which have a translational invariant Hamiltonian 
and where no quenched disorder is present.

This prejudice (on the need of quenched disorder( was so strong that it took a few years to realize 
that the replica method can also applied to system without quenched disorder.

There are many facts that clearly indicate the possibility of applying the replica methods to 
non-random systems.
\begin{itemize}
\item In the infinite range case there are pairs of systems with Hamiltonian respectively $H_{Q}$ and
$H$, where $H_{Q}$ contains quenched disorder and no disorder is present in $H$, such the high 
temperature expansion for the two systems coincide \cite{MPR,FH}.  It is natural to suppose that the 
free energies of the two system are identical at all temperatures, so that replica symmetry breaking 
can be applied to both.

\item It is possible to look for replica symmetry breaking in the expression for the free 
energy systems without disorder (e.g.  soft spheres) inside a given approximation (e.g.  
hypernetted chain) and find out that replica symmetry breaks at low temperature \cite{MP}.

\item In the replica method we can introduce coupled replica potentials \cite{PV} in order to
characterize the phase space of the system and these potentials can also be computed for non-random 
systems, obtaining the same results as for random systems \cite{FP}.  This may be done analytically 
for soft spheres using the same approximation as before \cite{CFP}.

\item It is now clear that the replica method may be applied any stochastic stable system.  Indeed
stochastic stable systems are the limit of disorder systems where the replica method can be 
applied without problems.  Systems without quenched disorder may be stochastically stable if the 
free energy is computed using the Cesareo limit (i.e.  averaging over $N$).
\end{itemize}

This new perspective allows us to use the replica method in systems quite different from the usual 
one, e.g.  structural glasses, where no quenched disorder is present.

\section{A DYNAMICAL APPROACH}
Although the predictions of the previous sections are quite clear, it is not so simple to test them 
for many reasons:
\begin{itemize}
\item They are valid  at thermal equilibrium,  a condition that is very difficult
to reach for this kind of systems, also in real experiments.
\item Experimentally is extremely difficult to measure the values of the microscopic variables, i.e 
all the spins of the system at a given moment.  These measurements can be done only in numerical 
simulations, where the observation time cannot be very large and only systems with less that $10^{4}$
degrees of freedom may be carried to thermal equilibrium.
\end{itemize}

A very important progress has been done when it was theoretically discovered that during the approach 
to equilibrium of the system, the fluctuation dissipation theorem is no more valid and the function 
$X(C)$, which describes the violations of the fluctuation dissipation theorem, (in some mean field 
models) is equal to the function $x(q)$ which is relevant for the statics
\cite{CUKU,FM}.  This equality is very interesting because function $X(C)$ can be measured 
relatively easily in off-equilibrium simulations
\cite{FRARIE}.

The temperature dependence of the function $X(C)$ (or equivalently $x(q)$) is interesting 
also because rather different systems can be classified in the same universality class according to 
the behaviour of this function.  It has been conjectured long time ago that the equilibrium 
properties of glasses are in the same universality class of some simple generalized spin glass 
models
\cite{KWT}.

Let us be more precise.  We concentrate our attention on a quantity $A(t)$.  We suppose that the 
system starts at time $t=0$ from an initial condition and subsequently it remains at a fixed 
temperature $T$.  If the initial configuration is at equilibrium at a temperature $T'>T$, we observe 
an off-equilibrium behaviour.  We can define a correlation function
\be
C(t,t_{w}) \equiv \lan A(t_{w}) A(t+t_{w})\ran
\ee
and the relaxation function
\be
G(t,t_{w}) \equiv \frac{ \de \lan A(t+t_{w})\ran}{\de \eps(t_{w})}{\Biggr |}_{\eps=0},
\ee
where we are considering the evolution in presence of a time dependent Hamiltonian in which we have
added the term
$ \int dt \eps(t) A(t) $.
 
The usual equilibrium fluctuation-dissipation theorem (FDT) tells us that
\be G^{eq}(t)= - \beta \frac{d C^{eq}(t)}{ dt}, \ee
where
\be
G^{eq}(t)=\lim_{t_w \to \infty} G(t,t_w), \ \ C^{eq}(t)=\lim_{t_w \to \infty} C(t,t_w).
\ee

It is convenient to define the relaxation function:
\be
R(t,t_{w})=\int_{0}^{t} d\tau G(t-\tau,t_{w}+\tau),\ \ R^{eq}(t)=\lim_{t_w \to \infty} R(t,t_w),
\ee
$R(t,t_{w})$ is the response of the system at time $t+t_{w}$ to a field acting for a time $t$ 
starting at $t_{w}$.  The usual FDT relation becomes
\be
R^{eq}(t)= \beta (C^{eq}(t)-C^{eq}(0)).
\ee

The off-equilibrium fluctuation-dissipation relation \cite{CUKU,FM} states that the response function 
and the correlation function satisfy the following relation for large $t_w$:
\be
R(t,t_w)\approx \beta \int_{C(t,t_w)}^{C(0,t_{w})}X(C) dC.  \label{OFDR}
\ee
If we plot $R$ versus $\beta C$ for large $t_{w}$ the data collapse on the same 
universal curve and the slope of that curve is $-X(C)$.  The function $X(C)$ is system dependent and 
its form tells us  interesting information. 
\begin{figure}[htbp]
\begin{center}
 \epsfxsize=450pt\leavevmode\epsffile{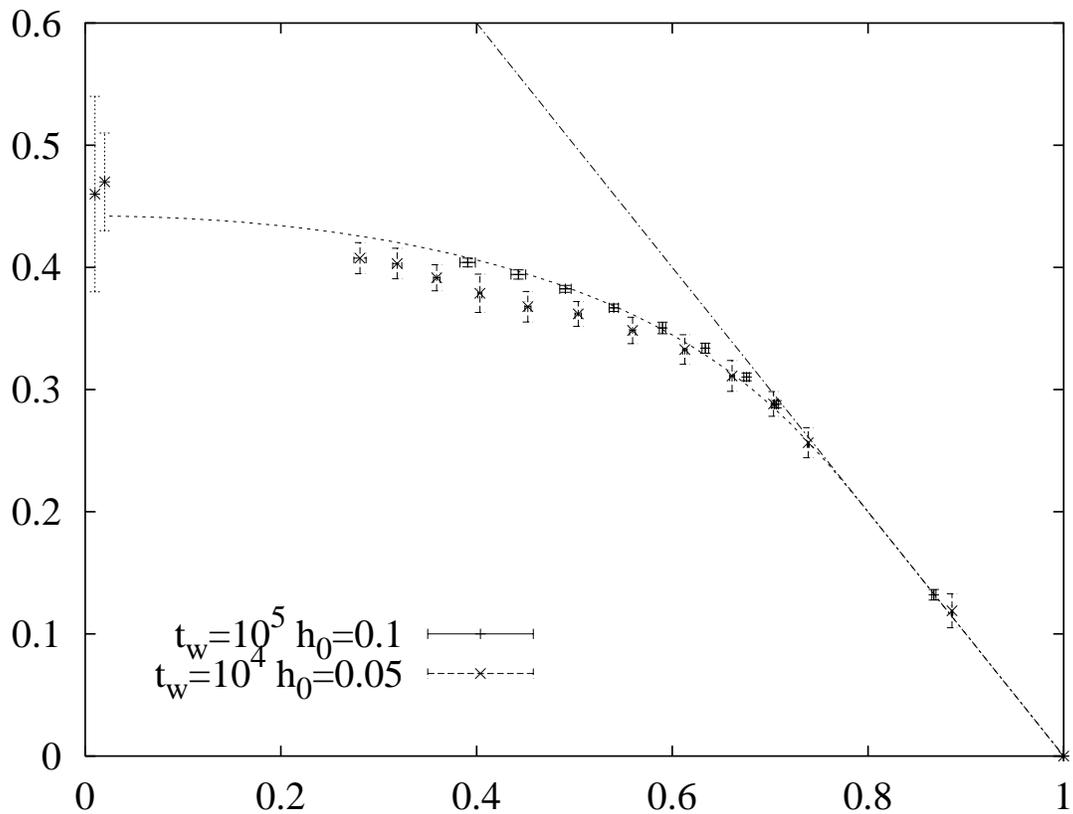}
\end{center}
\caption{The relaxation $ R$ times $T$ versus $ C$ at $T=0.7$ for the three
dimensional Ising spin glass [10].  The curve is the prediction for function $R(C)$ obtained from the 
equilibrium data.  The straight line is the FDT prediction.  We have plotted the data of the two 
runs: $t_w=10^5$, and $t_w=10^4$.}
\end{figure}

In the case of spin glasses this relation was shown to be valid in the mean field approximation, 
however there are quite general arguments that under the appropriate hypothesis it is also valid in 
general (also in short range models).  The proof is based on a dynamic version of stochastic 
stability: we must assume that in presence of a random perturbation (see eq.  (\ref{SSS}) the two 
limits ($t\to
\infty$ and $\eps
\to 0$) commute for the time dependent statistical expectation value of the appropriate quantities.

If we look more carefully to the graph of $R$ versus $\beta C$ we must distinguish two regions:
\begin{itemize}
\item A short time region where $X(C)=1$ (the so called FDT region) and $C$ belongs to the interval
$I$ 
(i.e. $C_1<C<C_2$.).

\item  A large time region (usually $t=O(t_w)$) where 
$C\notin I$ and $X(C)<1$.  In the same region the correlation function often satisfies an aging 
relation, i.e $C(t,t_w)$ depends only on the ration $s \equiv t/t_{w}$ in the region where both $t$ 
and $t_{w}$ are large: $C(t,t_w)\approx C^{a}(t/t_{w})$ \cite{B}.
\end{itemize}

In the simplest non trivial case, i.e.  one step replica symmetry breaking \cite{mpv,PAGE} , the 
function $X(C)$ is piecewise constant, i.e.
\be
X(C)= m \for C \in I,\ \ X(C)= 1 \for C \notin I \label{ONESTEP}.
\ee
One step replica symmetry breaking for glasses has been conjectured in ref.  \cite{KWT,PVAR}.

In all known cases in which one step replica symmetry holds, the quantity $m$ vanishes linearly with 
the temperature at small temperatures.  It often happens that $m=1$ at $T=T_{c}$ and $m(T)$ is 
roughly linear in the whole temperature range.  

Let us consider the case of spin glasses at zero magnetic field (in this case the replica symmetry 
is fully broken \cite{BOOK}).  The natural variable to consider is a single spin ($A=\si_{i})$).  
In this case the correlation $C(t,t_{w})$ is equal to the overlap among two configurations at time 
$t$ and $t_{w}$:
\be
C(t,t_{w})={\sum_{i=N}\si_{i}(t)\si_{i}(t_{w})\over N}.
\ee
The response function is just the magnetization in presence of an infinitesimal magnetic field.
In this case the situation is quite good because there are reliable simulations for the system at 
equilibrium \cite{BOOK}.

In fig.  (1) (taken from \cite {MPRR}) we plot the prediction for the function $R$ versus $C$, 
obtained at equilibrium (i.e.  using the equilibrium probability distribution of the overlaps, 
$P(q)$) by means of a simulation of a $16^3$ lattice using parallel 
tempering~\cite{HUKUNEMOTO,BOOK}.  The simulation involves the study of $900$ samples of a $ L=16$
lattice.

During the off-equilibrium simulations \cite {MPRR} in a first run without magnetic field the 
autocorrelation function has been computed.  In a second second run from $t=0$ until $t=t_w$ the 
magnetic field is zero and then (for $t \ge t_w$) there is an uniform magnetic field of small 
strength $h_0$.  The starting configurations were always chosen at random (i.e.  the system is 
suddenly quenched from $T=\infty$ to the simulation temperature $T$).

In fig.  (1) there are the results of the off-equilibrium simulations \cite {MPRR} where $t_w=10^5$ 
and $t_w=10^4$, with a maximum time of $5\cdot 10^6$ Monte Carlo sweeps.  The lattice size in was 
$64^{3}$, and $T=0.7$ (well inside the spin glass phase, the critical temperature is close to 1.0).  
We plot the response function $ R$ times $T$ (in this case $R$ is equal to $m/h_0$) against 
$C(t,t_w)$.  We have plotted also a straight line with slope $-1$ in order to control where the FDT 
is satisfied.  Finally we have plotted two points, in the left of the figure, that are obtained with 
the infinite time extrapolation of the magnetization.

The agreement among the absolute theoretical predictions (no free parameters) coming from the 
statics and the dynamical numerical data is quite remarkable.  These data show the correctness of the 
identification of the functions $x$ of the statics and $X$ of the dynamics.

\begin{figure}[htbp]
\begin{center}
 \epsfxsize=450pt\leavevmode\epsffile{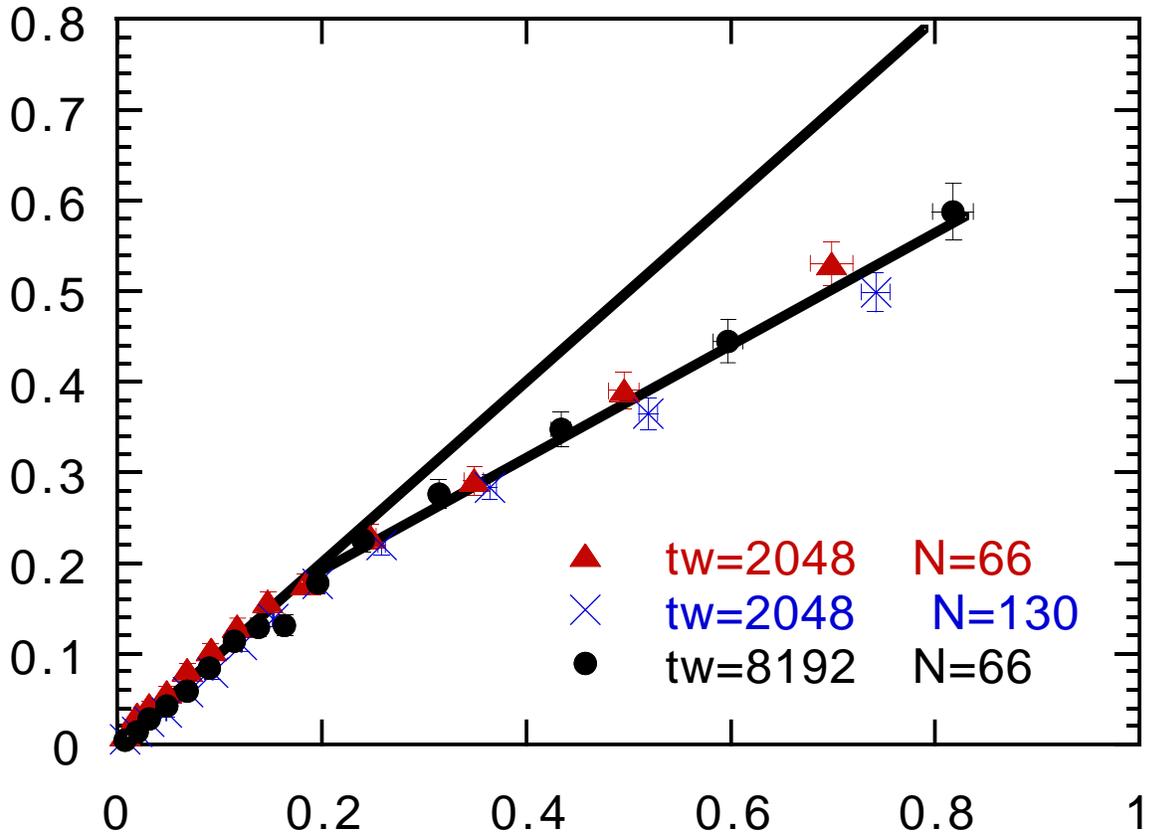}
\end{center}
\caption{ $R$ versus $\beta\Delta$ at $\Gamma=1.6$ for
$t_{w}=8192$ and $t_{w}=2048$ at $N=66$	and	for	$t_{w}=2048$ at	$N=130$.  The two straight lines
have slope 1 and .62 respectively.}
\end{figure}

It is quite interesting to note that numerical evaluation of the function $X(C)$ in glass forming
systems (i.e.  binary mixtures of soft spheres) strongly support the conjectures that glasses are 
systems in which the replica symmetry is broken at one step \cite{PAGE}.  Fig.  (2) shows the 
results for the function $R(\Delta)$ for 66 and 130 interacting particles.  Here the quantity 
$\Delta$ plays the same role of $1-C$ in spin glasses.The data for $R(\Delta)$ can be well fitted by 
two straight lines, as expected in the case of one step replica symmetry breaking

\section{CONCLUSIONS}

Replica theory for disordered systems provides a detailed pictures of the behaviour of glasses 
systems near and below the glass transition.  The theoretical predictions are in very good agreement 
with large scale numerical simulations \cite{BOOK,PAGE}.  Many of the unclear points (especially on 
the dynamics and on the approach to equilibrium) are now well understood.

The next step would be to test experimentally the core of the theory and to extract the function 
$x(q)$ from the data for the violation of the fluctuation dissipation theorem in off equilibrium 
dynamics.  The theoretical setting is well defined, we need however carefully planned experiments in 
order to measure the thermal noise correlations (the response is much easier).  I am confident that 
these experiments will be done in the next future.

\end{document}